\documentclass{aa}
\usepackage{graphics}
\begin{document}
\thesaurus{11(11.05.1; 11.05.02; 11.06.1; 13.09.1; 12.03.3;)}
\authorrunning{M. Stiavelli et al.\ }
\titlerunning{A candidate high-z elliptical galaxy}
%\singlespace
\title{VLT and HST observations of a candidate high redshift elliptical
galaxy in the Hubble Deep Field South
\thanks{Based on observations collected at the European
Southern Observatory, Paranal, Chile (VLT-UT1 Science Verification
Program) and with the NASA/ESA HST, obtained at the STScI, which is
operated by AURA, under NASA contract NAS5-26555. }}

\author{M. Stiavelli \inst{1,2,6}
\and T. Treu \inst{1,2} 
\and C.~M. Carollo \inst{3}\thanks{Hubble Fellow}
\and P. Rosati \inst{4}
\and R. Viezzer \inst{4}
\and S. Casertano \inst{1,6}
\and H. Ferguson \inst{1}
\and A. Fruchter \inst{1}
\and P. Madau \inst{1}
\and C. Martin \inst{1}
\and H. Teplitz \inst{5}
}

\offprints{M. Stiavelli}
\mail{M. Stiavelli} 

\institute{Space Telescope Science Institute, 3700 San Martin Dr.,
21218 MD, U.S.A.; mstiavel@stsci.edu, treu@stsci.edu \and Scuola
Normale Superiore, Piazza dei Cavalieri 7, I56126, Pisa, Italy;
mstiavel@astro.sns.it, treu@cibs.sns.it \and Johns Hopkins University,
3701 San Martin Dr., 21218 MD, U.S.A.; marci@pha.jhu.edu 
\and European Southern Observatory, Karl-Schwarzschild Str. 2, D85748,
Garching; prosati@eso.org 
\and Goddard Space Flight Center, Greenbelt, MD 20771, U.S.A.;
hit@binary.gsfc.nasa.gov
\and On assignment from the Space Science Department of the European
Space Agency} \date{Received / Accepted}

\maketitle

\begin{abstract}
The combined use of the ESO Very Large Telescope (VLT) UT1 Science
Verification (SV) images and of the Hubble Space Telescope (HST)
Hubble Deep Field South observations allows us to strengthen the
identification as a candidate elliptical galaxy of the Extremely Red
Object HDFS 223251-603910 previously identified by us on the basis of
NICMOS and Cerro Tololo Interamerican Observatory imaging.  The
photometry presented here includes VLT data in U, B, V, R, I, a STIS
unfiltered image, NICMOS J, H, and K band data, thus combining the
16.5 hours of VLT SV exposures with 101 hours of HST observing.
The object is detected in all images except the
VLT U band and is one of the reddest known with B-K$=9.7\pm0.5$.  We
consider a wide range of models with different ages, metallicities,
star formation histories and dust content, and conclude that the
observed spectral energy distribution agrees best with that of an old
elliptical galaxy at redshift just below 2. Alternative possibilities
are discussed in light of their likelihood and of the perspective of
spectroscopic confirmation.


\end{abstract}

\section{Introduction}

\label{sec:intro}

The formation process of elliptical galaxies is still unknown. Direct
observations of elliptical galaxies show that ellipticals exist up to
a redshift of one, but the statistics are too poor to draw firm
conclusions concerning the evolution of the population as a whole
(see, e.g., Kauffmann et al., 1996; Im et al. 1996; and Im and
Casertano 1998).  Indirect evidence based on the evolution with
redshift of the Fundamental Plane of elliptical galaxies (van Dokkum
et al. 1998), and on the global correlation of metallicity with
velocity dispersion (Ziegler \& Bender et al. 1997), as well as direct
measurement of Balmer line strengths (see, e.g., Bressan et al. 1996),
indicate that the stellar populations of ellipticals are old, with
ages in excess of 10 Gyrs. Unfortunately, the uncertainties in the
basic cosmological parameters do not allow us to firmly convert age
estimates into a formation redshift. In particular, if the geometry of
our Universe is dominated by a cosmological constant, stars in
ellipticals could have formed at a redshift of 1.65 and still be older
than 10 Gyrs at the present time (with $\Omega=0.35$,
$\Omega_\Lambda=0.65$, $H_0=65$ km s$^{-1}$ Mpc$^{-1}$).

The strong correlation between dynamics and stellar population
properties argues for a tightly defined formation process.  On the
other hand, models in which elliptical galaxies form as the result of
stellar-gaseous mergers of disk systems at moderate redshifts appear
to reproduce successfully the observed properties of elliptical
galaxies (see, e.g., Kauffmann and Charlot 1998 and references
therein); in such models, most stars form before the galaxy itself is
finally assembled.  However, the predictive power of these models is
limited by our modest knowledge on star formation processes. In
contrast with their sophisticated treatment of the evolution of
gravitational clustering, typical galaxy formation and evolution codes
compute star formation according to simple empirical relations which
may not be representative of realistic star formation processes.  In
fact, on the basis of entirely different theoretical arguments
including a detailed description of ISM properties but no cosmology
nor dynamical evolution, one could instead argue that spheroidal
systems are preferentially formed at high redshift (Spaans \& Carollo
1997, Carollo et al. 1997a). Thus, it appears that the present
stalemate between different theories and low-to-moderate redshift
observations awaits an observational answer.

Important clues may indeed be provided by the discovery of a
population of Extremely Red Objects (EROs): relatively bright in the
NIR (K$\sim 19$--$20$) and very faint in the optical ((R$-$K)$>5$
typically \footnote{The definition of EROs is not uniform in the
literature. Other values are, for example, (R$-$K)$>4.5$ (Elston et
al.\, 1991) and (R$-$K)$>6$ (Graham \& Dey, 1996).}). As was
immediately suggested, such a red color can be produced by an ``old''
stellar population at high redshift (e. g., LBDS 53W091 at z=1.55,
\cite{SP}) or by a younger stellar population enshrouded by dust lanes
(e. g., HR10 at z=1.44, \cite{GD}, with one detected emission
line). These objects have generally subarcsecond sizes and therefore
very little morphological information can be gathered with ground
based telescopes. Nevertheless, in the case of HR10, the morphology
appeared to be irregular, thus strengthening the case for the object
being a late-type or irregular galaxy. Furthermore, the presence of a
young stellar population was suggested also by the relatively blue B-I
color of HR10 (B-I$\approx1.7$).

Besides the two prototypical (and particularly bright) cases of LBDS
53W091 and HR10 where it was possible to determine the redshift using
the Keck 10-m telescope, the ``dust-age'' degeneracy in the broad band
colors is generally coupled with the redshift and metallicity
uncertainties. In other words, the extremely red R-K observed in
the EROs can be explained by different combinations of star formation
history, redshift, dust content and chemical composition. The only way
to photometrically reduce this degeneracy is to obtain multiple broad
band colors, over the widest possible spectral range, and to measure
morphological properties.

In this Letter we present photometric information of unprecedented
completeness about the extremely red $R^{1/4}$ galaxy identified by
Treu et al. (1998a) in the Test Image of the Hubble Deep Field
South. Due to its particular location the galaxy has been observed
with the VLT-UT1 and the Hubble Space Telescope (HST), for a total of
9 bands covering the spectral bands U through K. The VLT was able to
detect it in the optical down to B=$29\pm0.5$ and to provide us with
the optical SED, while HST gives us IR colors and multiband
morphological information.

The data strengthen the early identification as an optimal candidate high
redshift elliptical galaxy. If spectroscopically confirmed, this
identification would imply that some ellipticals can be in place already
at redshift 2 and that their stars have formed at even higher redshifts.
In Section 2 and the photometric measurements. In Section 3
we discuss the identification of the object. 
In this Letter the Hubble constant is assumed to
be 100 $h$ km/s/Mpc, with $h=0.65$ where needed.

\section{Observations}

The HDFS NICMOS field was observed with the Test Camera on VLT-UT1
during the VLT Science Verification (Renzini, 1998) and with NICMOS on
board the Hubble Space Telescope during the HDFS (Williams et
al. 1999) campaign. The field was observed for an additional 9 orbits
with the Space Telescope Imaging Spectrograph (STIS) to obtain a
broad, high resolution, visible image, and, as one of the flanking
fields, with the Wide Field and Planetary Camera 2
(WFPC2). Unfortunately, we were not able to include the WFPC2 images
in our analysis as, at the time of this writing, the final reduced
frame had not been released. We refer to Fruchter et al.  (1999) for
the description of the observations and the data reduction.

In the following discussion we shall indicate the Hubble Space
Telscope filters in the following way: C = STIS 50CCD (clear);
J$_{110}$= NICMOS F110W; H$_{160}$= NICMOS F160W; K$_{222}$= NICMOS
F222M.

\subsection{Morphology}

\label{sec:morfe}

\begin{figure}
\resizebox{\hsize}{!}{\includegraphics{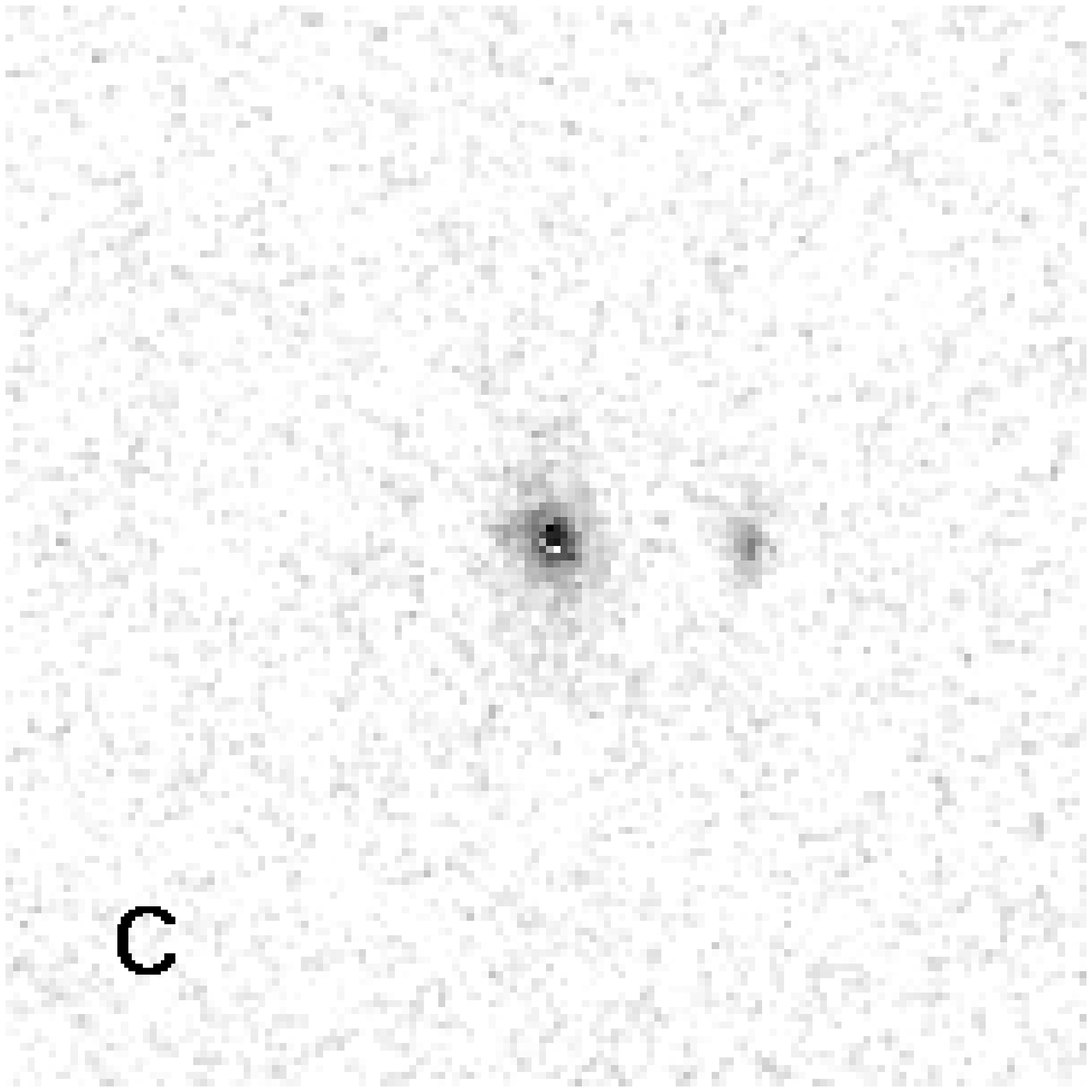}\includegraphics{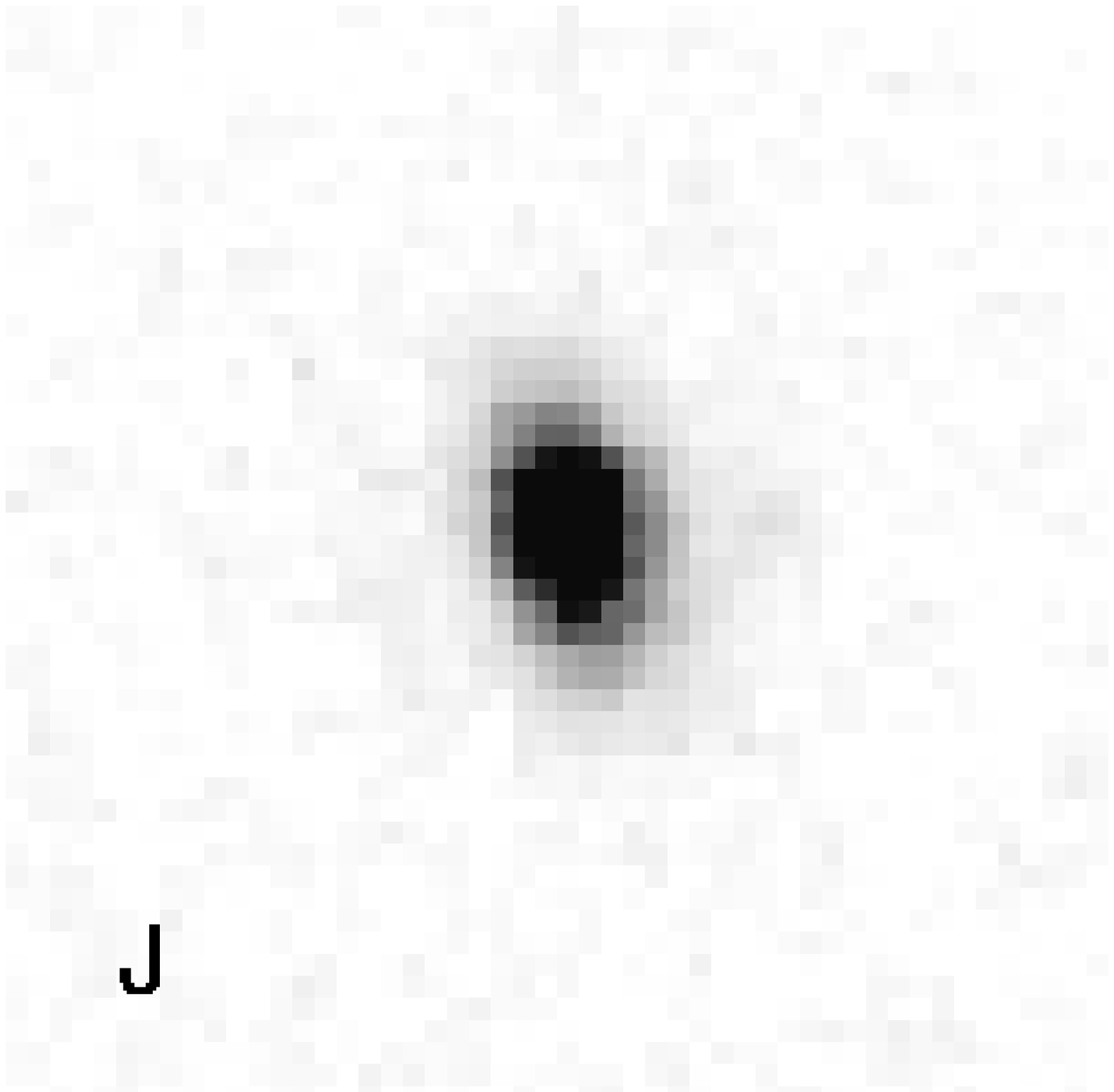}\includegraphics{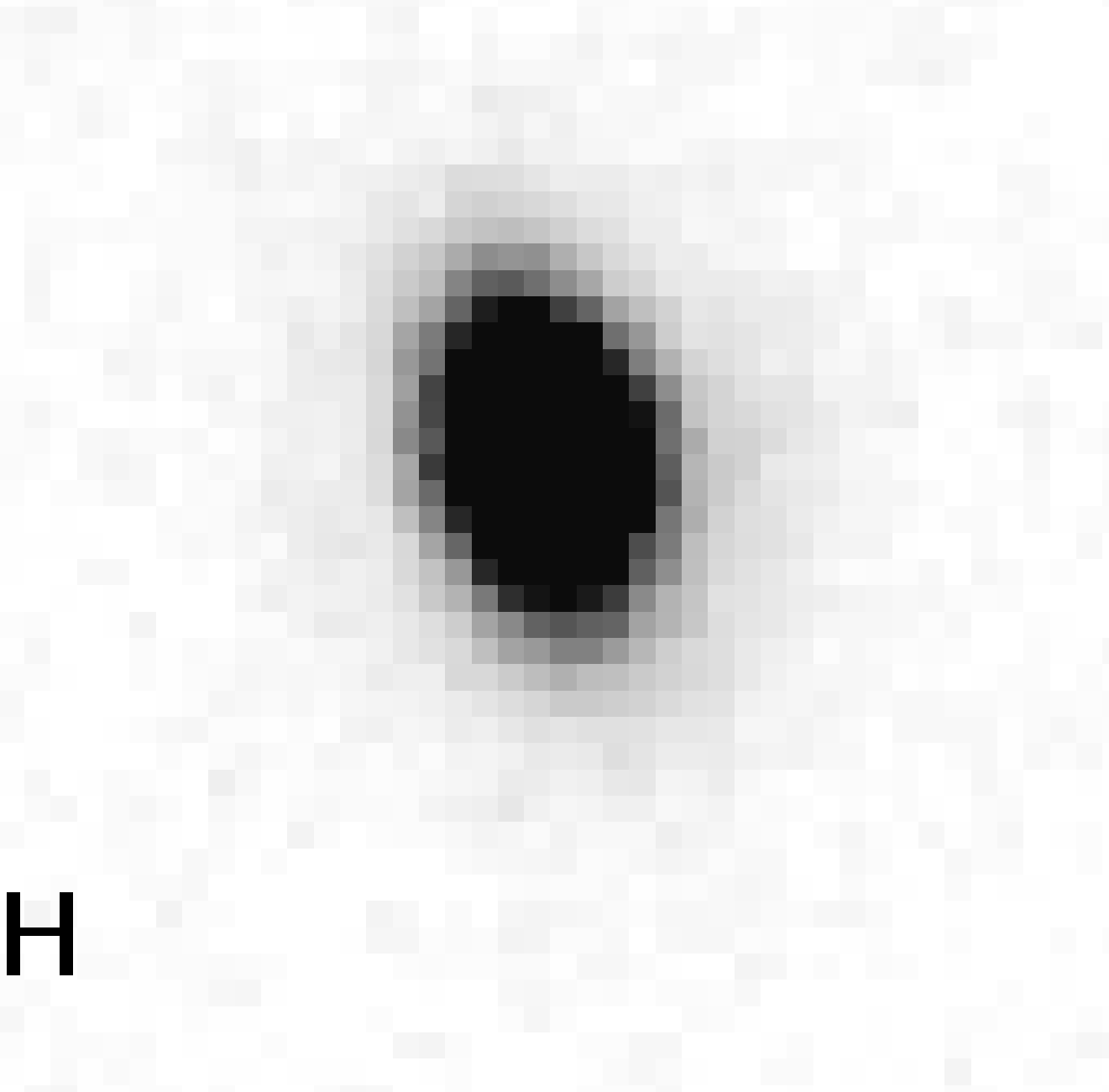}\includegraphics{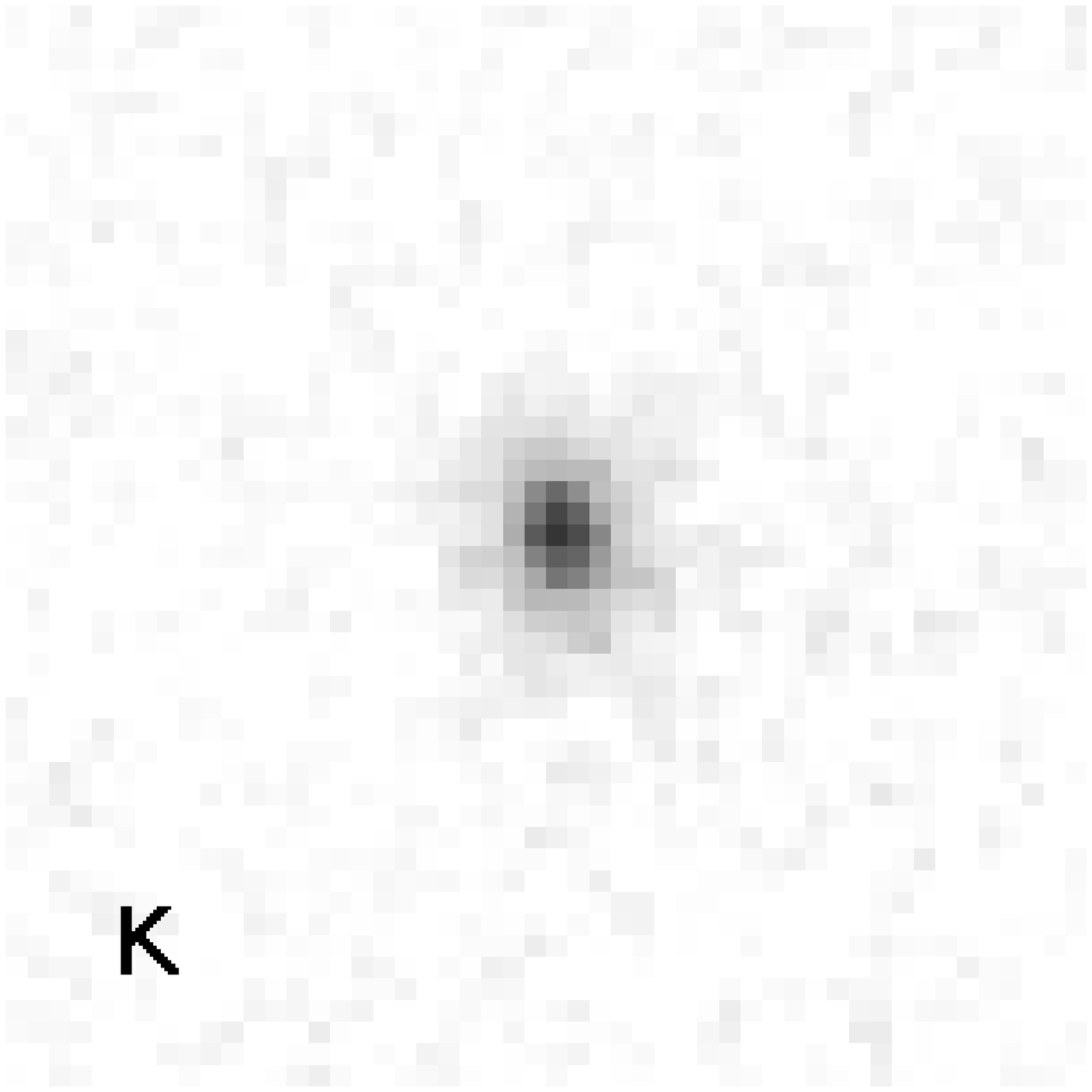}}
\resizebox{\hsize}{!}{\includegraphics{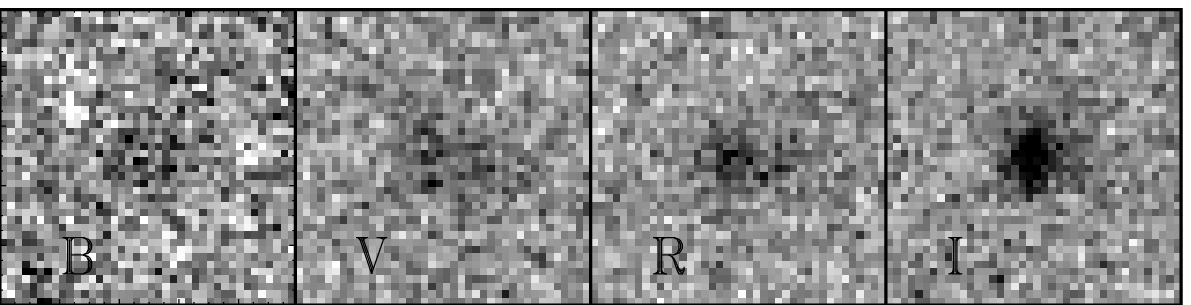}}
\caption{HST and VLT images of HDFS 223251-603910 {\bf C.} STIS clear
.  {\bf J.} NICMOS J$_{110}$ {\bf H.} NICMOS H$_{160}$ {\bf K.} NICMOS
K$_{222}$.  {\bf B.} VLT B {\bf V.} VLT V {\bf R.} VLT R {\bf I.} VLT
I.  The images are $3\farcs75$ on a side. A faint blue source very
near to the center of the galaxy is visible in the STIS and NICMOS J
image.  }
\label{fig:IM}
\end{figure}

The galaxy is clearly resolved in the NICMOS and STIS images, as shown
in Figure~\ref{fig:IM}. To strenghten the visual classification as an
early type galaxy, we fitted the luminosity profile with an $R^{1/4}$
model and an exponential model. We performed the fit independently on
each STIS and NICMOS image using two different techniques: the
two-dimensional fit code described in Treu et al. (1998b) and the
luminosity profile code described in Carollo et al.\ (1997b). With the
latter technique we fitted also a bulge+disk model. We used in each
case two different stars as PSFs, in order to check the sensitivity of
our results to a particular choice of PSF. We confirm the Treu et al.\
(1998a) result that only an $R^{1/4}$ profile provides an acceptable
fit.  The high resolution STIS observations confirm the size we
determined from the earlier NICMOS observations; note that the derived
half-light radius is larger than both the STIS pixel size and PSF
FWHM, making our result more robust. The uncertainties listed in Table
\ref{tab:data} include the error on the sky subtraction and the error
due to PSF estimates.  We cannot exclude the presence of a faint,
extended disk, but we have verified that our results are not
significantly affected by the presence of such disk components by
carrying out a variety of bulge+disk fitting experiments (e.g., we
find that the ``bulge'' magnitude does not change by more than $\sim$
0.1 mag).

\begin{figure}
\resizebox{\hsize}{!}{\includegraphics{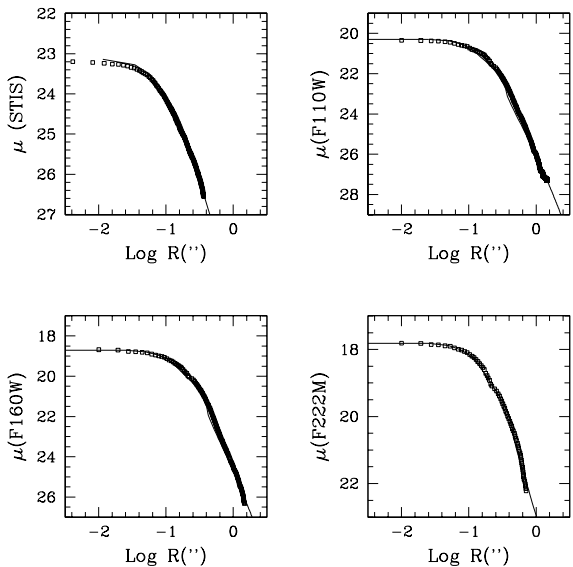}}

\caption{Luminosity profile of the galaxy (squares) in the STIS and
NICMOS passbands. The PSF-convolved best-fit $R^{1/4}\,$ profile is
shown for comparison (solid line). }
\label{fig:prof}
\end{figure}

\subsection{Colors}
\label{ssec:col}

The multiband colors of HDFS 223251-603910 are summarized in Table
\ref{tab:data}. The STIS and NICMOS magnitudes are derived by fitting
$R^{1/4}$ luminosity profiles as described in the previous section but
were also checked with aperture measurements and found to be
consistent to within $\sim$0.1 mags. For the VLT data we give aperture
magnitudes corrected for PSF losses. We estimate the systematic error
in the correction to be in the range $0.05-0.1$ mags. The
uncertainties listed ($\delta$m) do not include systematic calibration
errors, which might be as high as 0.1 mags for the NICMOS magnitudes
and as high as 0.05 mags for the VLT magnitudes. We adopt here the
NICMOS Pipeline zeropoints. Particularly the H$_{160}$ zeropoint used
here is 0.08 mags higher than the one used in Treu et al. (1998a). The
VLT zero points in the final coadded images, as released on the WEB
site http://www.hq.eso.org/paranal/sv/, were calculated by matching
the photometry in the final stacks with that of a single photometric
night.  The overall consistency of the zeropoints was checked by
comparing the VLT data with CTIO data and the main WFPC2 HDFS
field. The colors listed in Table 1 are corrected for galactic
extinction, using the value E(B-V)=0.026 reported by Schlegel et
al. (1998).

\section{Identification and conclusions}

\begin{figure}
\resizebox{\hsize}{!}{\includegraphics{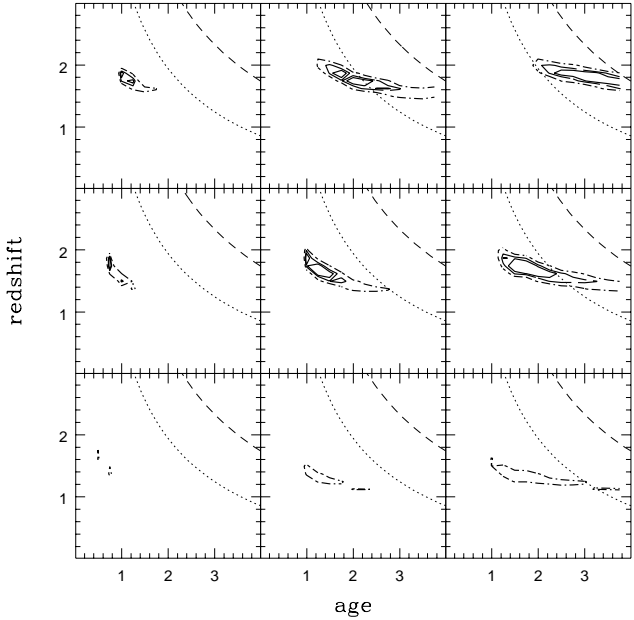}}
\caption{Contour plots of the $\chi^2$. The colors of a grid of models
(see text for details) have been computed and compared with the
observed ones. The metallicity of the stellar populations is
Z/Z$_{\odot}$=1,0.4,0.2 from left to right, the dust reddening is
A$_V$=0,0.5,1 from top to bottom. The contours correspond to a
probability of 68, 80 and 95 \% from the inside to the outside.  The
two curves in the upper left corner represent the age of the Universe
as a function of redshift for two different cosmologies (see
text). The object must be younger than the Universe and therefore only
the region below the curves is allowed. Its redshift is better
constrained than its age. }

\label{fig:chi2}
\end{figure}

\begin{figure}
\resizebox{\hsize}{!}{\includegraphics{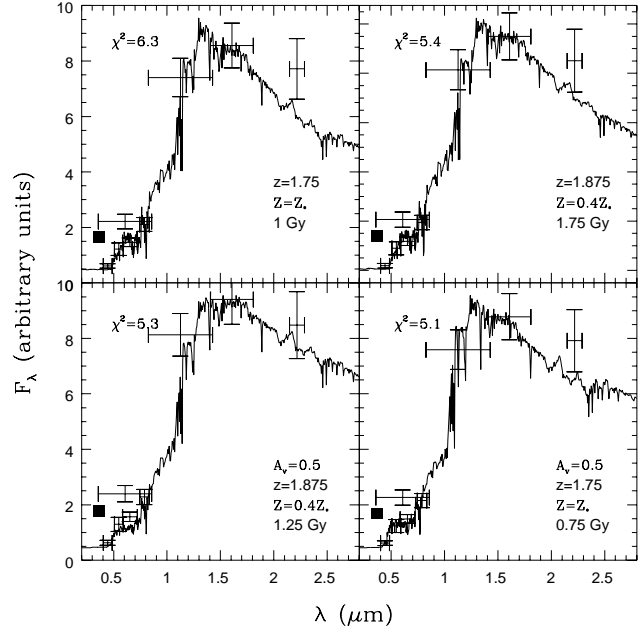}}
\caption{The best fitting spectra for the cases with
Z=1,0.4Z$_{\odot}$ and $A_{V}=0,0.5$ are shown, togheter with the
observed photometric measurements and the.  The filled square is the U band
VLT upper limit. The error bars on the wavelength axis represent the
filter bandwidths. The main contribuition to STIS CCD clear filter
comes from the reddest part of the spectrum. The $\chi^2$ with seven degrees of freedom is shown in the upper left corner.}
\label{fig:spectra}
\end{figure}

Using nine broad band magnitudes derived from VLT and HST data, we are
now able to constrain the SED of ERO HDFS 223251-603910 very well. In
order to provide a firm identification, we computed the colors of a
large grid of models, with different redshift, age, metallicity and
dust content, using the population synhtesis code by Bruzual \&
Charlot (1993, GISSEL96 version).  At this stage, to avoid
further degeneracies, we limited our analysis to single burst stellar
population models. Different star formation histories with the same
age and chemical composition, would produce intrinsically bluer
colors. Specifically we considered models with metallicity
Z/Z$_{\odot}$=1,0.4,0.2 and dust reddening A$_V$=0,0.5,1 (using the
extinction law from Cardelli et al. 1989).  Then we computed the
$\chi^2$ of the synthetic colors with the observed ones. In
Figure~\ref{fig:chi2} the contour levels of probability 68, 80 and
95 \% are shown.  In the nine panel the metallicity decreases from
left to right, while the dust reddening increases from top to
bottom. As can be seen, the redshift is relatively well constrained,
but the age of the object is much more uncertain. This is a
consequence of the so-called age-metallicity degeneracy (e.g., Worthey,
1994), i.e., that broad band colors cannot distinguish the effects of
age and metallicity. Only spectroscopy could constrain both the
metal content and the age of the object.  The dotted and dashed lines
on the upper left corner of the plots are the age of Universe as a
function of redshift (h=0.65; dotted : $\Omega=1,\Omega_{\Lambda}=0$,
dashed : $\Omega=0.3,\Omega_{\Lambda}=0.7$). As the galaxy cannot be
older than the Universe, only the region below the curves is allowed.
The best fitting models of our grid in the cases A$_v=0,0.5$ and
Z/Z$_{\odot}$=1,0.4 are shown in Figure \ref{fig:spectra}, together
with the photometric points. The $\chi^2$ and the model
characteristics are also shown in the figure.

As a further check we also compared the colors of HDFS 223251-603910
with the colors of highly reddended starbust galaxies (\cite{D}, and
references therein) and we did not find any combination that could
reproduce the measured Spectral Energy Distribution.

As already noticed in Treu et al. (1998a), the luminosity of this
galaxy is consistent with its identification as an elliptical galaxy
at $z\simeq 1.8$, e.g., for $\Omega=0.35$ and $\Omega_\Lambda=0.65$,
the galaxy would have, after passive dimming down to z=0, M$_V=-21.4$
and an effective radius of about 4 kpc.

Thus, all the photometric data collected so far are consistent with
the identification of this object as a high redshift elliptical. This
renders HDFS 223251-603910 a prime target for spectroscopic followup.
This is within reach with IR spectrograph on 8m class telscopes, such
as ISAAC on the VLT. If confirmed, its identification will demonstrate
that ellipticals can already be assembled at redshift almost
2. Furthermore, by measuring its absorption line strengths we will
determine in great detail its physical properties, such as age, metal
content and star formation history, thus improving our knowledge of
the early stages of galaxy formation.

\begin{table}
\caption{Photometric properties. For every passband total magnitudes
are listed with their errors, except for U and in which case the 95\%
CL limits is shown. The errors on the magnitudes do not include
systematic calibration errors (see Section \ref{ssec:col}). The
effective radius, as measured by fitting the $R^{1/4}$ profile to the
HST images is also listed. The exposure times (texp), the zero points
(zp) and colour terms (ct), referred to the ESO and STScI released
data, are shown for reference. }

\label{tab:data}
{\scriptsize
\begin{tabular}{c c c c c c}
Band & m $\pm \delta$ m & $r_{\textrm{e}}$ $\pm \delta$$r_{\textrm{e}}$ &
texp (s) & zp & ct \\
\hline
U               & $<26.5$      &  -             &  17788  & 31.39 & 0.239(U-B)\\    
B               & 29$\pm$0.5   &  -            &  10200   & 33.26 & 0.128(B-V)\\    
V               & 26.7$\pm$0.3  &  -            &  14400  & 33.87 & -0.066(B-V)\\
R               & 25.81$\pm$0.15 &  -           &  7200   & 34.13&  -0.061(V-R)\\
C               & 25.38$\pm$0.15 & 0$\farcs26$$\pm$0$\farcs$06 &  25900 & 26.15& -\\
I               & 24.65$\pm$0.15 &  -  &  10158           & 33.55 & 0.023(V-I)\\
J$_{110}$       & 22.01$\pm$0.10 & 0$\farcs17$$\pm$0$\farcs$04     &  108539 & 22.16 & -\\
H$_{160}$       & 20.47$\pm$0.10 & 0$\farcs16$$\pm$0$\farcs$04     &  128441 & 21.52 & -\\
K$_{222}$       & 19.33$\pm$0.15 & 0$\farcs20$$\pm$0$\farcs$05   &  103163   & 20.00 & -\\
\hline  
\end{tabular}}
\end{table}

\begin{acknowledgements}

The authors would like to thank Bob Williams, Alvio Renzini, the
VLT-UT1 Science Verification Team, and the Hubble Deep Field South
Team, for having provided the astronomical community with this
exceptional data set.
\label{sec:ak}

\end{acknowledgements}


\begin{thebibliography}{}

%\bibitem[Baugh et al.\ 1998]{CF} Baugh C., Cole S., Frenk C.,
%Lacey C. 1998, ApJ, 498, 504

%\bibitem[Bertin \& Arnout 1996]{SEx} Bertin E., Arnouts S., 1996, A\&AS, 117, 393 

\bibitem[Bressan et al.\ 1996]{Bressan} Bressan A., Chiosi C., Tantalo R.,
1996, A\&A, 311, 425 

\bibitem[Bruzual \& Charlot 1993]{BC93} Bruzual A.~G., Charlot S.
1993, ApJ, 405, 538

\bibitem[Calzetti 1998 ]{D} Calzetti D., 1998, astro-ph/9806083

\bibitem[Cardelli et al.\ 1989]{dust} Cardelli J., Clayton G., 
Mathis J. 1989, ApJ, 345, 245

\bibitem[Carollo et al.\ 1997a]{Eform} Carollo C.~M., Spaans M., 
Stiavelli M., Mihos J.~C., 1997, in ``Science with the NGST'', E.P. Smith
and A. Koratkar eds, p. 237.

\bibitem[Carollo et al.\ 1997b]{MCetal} Carollo C.~M., Franx M.,
Illingworth G.~D., Forbes D.~A. 1997b, ApJ, 481, 710

%\bibitem[Casertano et al.\ 1999]{c99} Casertano S. et al. 1999, in preparation

%\bibitem[Cimatti et al.\ 1998]{Cima} Cimatti A., Andreani P., R\"ottgering H., Tilanus R. 1998, Nature, 392, 895

%\bibitem[Dey et al.\ 1995]{DSD} Dey A., Spinrad H., Dickinson M.~E.
%1995, ApJ, 440, 515

%\bibitem[Dickinson 1998]{Dick} Dickinson M.~E., 1998, preprint
%(astro-ph/9802064v2)

%\bibitem[Driver et al.\ 1998]{EE} Driver S.~P., Fern\'andez-Soto A.,
%Couch W.~J., et al.\
%Odewhan S.~C., Windhorst S.~C., Phillipps S., Lanzetta K., Yahil A. 
%1998, ApJ, 496, L93 

%\bibitem[Elston et al.\ 1988]{E} Elston R., Rieke G.~H., Rieke
%M.~J. 1988, ApJ, 331, L77

\bibitem[Elston et al.\ 1991]{E2} Elston R., Rieke G.~H., Rieke
M.~J. 1991, in Elston (ed.):``Astrophysics with Infrared Arrays'',
ASP, San Francisco.

%\bibitem[Fruchter \& Hook 1998]{drizzle} Fruchter A.~S., Hook
%R.~N. 1998,  submitted to PASP

\bibitem[Fruchter et al. 1999]{F99} Fruchter A.~S., 1999, in preparation

%\bibitem[Francis et al.\ 1996]{z238} Francis P.~J., Woodgate B.~E.,
%Warren S.~J., et al.\ 1996, ApJ, 457, 490

\bibitem[Graham \& Dey 1996]{GD} Graham J.~R., Dey A. 1996, ApJ, 471, 720

%\bibitem[Hu \& Ridgway 1994]{Hu} Hu E.~M., Ridgway S.~E. 1994, AJ, 107, 1303

\bibitem[Im et al.\ 1996]{z1noevol} Im M., Griffiths R., Ratnatunga K.U.,
Sarajedini V.L., 1996, ApJ, 461, L79

\bibitem[Im and Casertano 1998]{z1fight} Im M., Casertano S., 1998,
preprint.

\bibitem[Kauffmann and Charlot 1998]{colormag} Kauffmann G., Charlot S.,
1998, MNRAS, 294, 705             

\bibitem[Kauffmann et al.\ 1996]{z1evol} Kauffmann G., Charlot S., White
S.D.M., 1996, MNRAS, 283, L117

%\bibitem[Landolt 1992]{Lan} Landolt A.~U. 1992, AJ, 104, 340

%\bibitem[Lilly et al.\ 1995]{CFRS} Lilly S.~J., Tresse L., Hammer F., et al.\ 
%Crampton D., Le F\'evre O. 
%1995, ApJ, 455, 108

%\bibitem[Lowenthal et al.\ 1997]{DEEP} Lowenthal J.~D., Koo J.~D.,
%Guzman R., et al.\ 1997, ApJ, 481, 673

%\bibitem[Lucas et al.\ 1999]{L99} Lucas R.~A., 1999, in preparation

%\bibitem[Madau et al.\ 1996]{MP} Madau P., Ferguson H.~C., Dickinson,
%M.~E. et al.\
%Giavalisco M., Steidel C.~C., Fruchter A. 
%1996, MNRAS, 283, 1388

%\bibitem[McCarthy et al.\ 1998]{Mc} McCarthy P., Yan L.,
%Storrie-Lombardi L., Weymann R.~J., 1998, in: NICMOS and the VLT, ESO
%Conference Workshop Proceedings 55, W. Freudling and Richard Hook eds.

%\bibitem[Maoz 1997]{Mao} Maoz D. 1997, ApJ, 490, 135

%\bibitem[M{\o}ller et al.\ 1995]{MSZ} M{\o}ller P., Stiavelli 
%M., Zeilinger W. 1995, MNRAS, 276, 979

%\bibitem[Moustakas et al.\ 1997]{Mou} Moustakas L.~A., Davis M.,
%Graham J.~R., et al.\
% Silk J., Peterson B.~A., Yoshii Y. 
%1997, ApJ, 475, 445

\bibitem[Renzini 1998]{Renzini} Renzini, 1998, ESO Messenger, 93, 1 

%\bibitem[Schade et al.\ 1997]{K97} Schade D., Barrientos L.,
%L\'opez-Cruz O. 1997, ApJ, 477, L17

\bibitem[Schlegel et al.\ 1998]{EXMAP} Schlegel D.J. et al.\ 1998,
  ApJ, 500, 525

%\bibitem[Soifer et al.\ 1998]{So98} Soifer B.~T., Neugebauer G.,
%Franx M., et al.\
%Matthews K., Illingworth G.~D.
%1998, ApJ, 501, L171

\bibitem[Spaans \& Carollo 1997]{SC} Spaans M., Carollo C.M., 1997,
  ApJ, 482, 93

\bibitem[Spinrad et al.\ 1997]{SP} Spinrad H., Dey A., Stern D., et al.\
%Dunlop J., Peacock J., Jimenez R., Windhorst R. 
1997, ApJ, 484, 581

%\bibitem[Stanford et al.\ 1997]{Sta} Stanford S.~A., Elston R.,
%Eisenhardt P.~R., et al.\
%%Spinrad H., Stern D., Dey A. 
%1997, AJ, 114, 2232

%\bibitem[Steidel et al.\ 1996]{S96} Steidel C.~C., Giavalisco M., 
%Dickinson M.~E., Adelberger K.~L., 1996 ApJ, 462, L17

%\bibitem[Steidel et al.\ 1998]{S98} Steidel C.~C., Adelberger K.~L.,
%Dickinson M.~E., et al.\
%Giavalisco M., Pettini M., Kellogg M. 
%1998, ApJ, 492, 428


\bibitem[Treu et al.1998a]{ERO-I} Treu T., Stiavelli M., Walker A. R.,
et al.\ 1998a, AA, 340, L10.

\bibitem[Treu et al.\ 1998b]{NOI} Treu T., Stiavelli M.,
Casertano S., et al.\
%M{\o}ller P., Bertin G. 
1998b, submitted to MNRAS


\bibitem[van Dokkum et al.\ 1998]{DFKI98} van Dokkum P., Franx M.,
Kelson D., Illingworth, G.~D. 1998, ApJL, 504, 17

%\bibitem[Williams et al.\ 1996]{HDF} Williams R.~E. et al.\  1996, AJ, 
%112, 1335

%\bibitem[Williams et al.\ 1997]{HDFS} Williams R.~E. et al.\ 1997,
%AAS, \#191.8508

\bibitem[Williams et al.\ 1997]{HDFS} Williams R.~E. et al.\, 1999, in
  preparation

\bibitem[Worthey 1994]{W94} Worthey G., 1994, ApJS, 95, 107

\bibitem[Ziegler and Bender 1997]{mgsig} Ziegler B.L., Bender R.,
1997, MNRAS, 291, 527

%\bibitem[Zepf 1997]{Zepf} Zepf S., 1997, Nature, 390, 377

\end{thebibliography}
\end{document}